\documentclass[conference,a4paper]{IEEEtran}

\usepackage{algorithm,makecell,algorithmic,amsbsy,amsmath,amssymb,epsfig,bbm,mathrsfs,multirow,amsthm}
\usepackage{array,multirow,graphicx}
\usepackage[english]{babel}
\usepackage[justification=centering]{caption}
\usepackage{url}
\usepackage{threeparttable}
\usepackage{epstopdf}
\usepackage{subfigure}

\begin{document}

\title{Joint Service Pricing and Cooperative Relay Communication for Federated Learning}

\author{
Shaohan Feng$^1$, Dusit Niyato$^1$, Ping Wang$^2$, Dong In Kim$^3$, and Ying-Chang Liang$^4$\\
$^1$ School of Computer Engineering, Nanyang Technological University, Singapore	\\
$^2$ Department of Electrical Engineering and Computer Science, York University, Canada \\
$^3$ School of Information and Communication Engineering, Sungkyunkwan University, Korea\\
$^4$ CINC, University of Electronic Science and Technology of China, China\vspace{-5mm}	}

\maketitle

\begin{abstract}
For the sake of protecting data privacy and due to the rapid development of mobile devices, e.g., powerful central processing unit (CPU) and nascent neural processing unit (NPU), collaborative machine learning on mobile devices, e.g., federated learning, has been envisioned as a new AI approach with broad application prospects. However, the learning process of the existing federated learning platforms rely on the direct communication between the model owner, e.g., central cloud or edge server, and the mobile devices for transferring the model update. Such a direct communication may be energy inefficient or even unavailable in mobile environments. In this paper, we consider adopting the relay network to construct a cooperative communication platform for supporting model update transfer and trading. In the system, the mobile devices generate model updates based on their training data. The model updates are then forwarded to the model owner through the cooperative relay network. The model owner enjoys the learning service provided by the mobile devices. In return, the mobile devices charge the model owner certain prices. Due to the coupled interference of wireless transmission among the mobile devices that use the same relay node, the rational mobile devices have to choose their relay nodes as well as deciding on their transmission powers. Thus, we formulate a Stackelberg game model to investigate the interaction among the mobile devices and that between the mobile devices and the model owner. The Stackelberg equilibrium is investigated by capitalizing on the exterior point method. Moreover, we provide a series of insightful analytical and numerical results on the equilibrium of the Stackelberg game.
\end{abstract}


\section{Introduction}
\label{sec:introduction}

Due to the explosive growth of smart IoT devices at the edge of the Internet, massive data collection through embedded sensors on mobile devices, e.g., crowdsensing, for machine learning has found a number of applications and gained tremendous popularity rapidly~\cite{feng2018competitive}. However, the existing machine learning approaches rely on centralized storage of the training data. Consequently, they usually face a series of data security and privacy issues, e.g., data abuse and information leakage. A recent report from Ponemon Institute suggests an average cost of over \$200 for per record of data breach~\cite{opderbeck2015cybersecurity}. Such a high economic loss has hindered the adoption of data sharing among the different entities, and the machine learning that requiring centralized data storage is facing great challenges.

To overcome the limitations of traditional machine learning in the protection of data privacy, a novel paradigm has been proposed. In~\cite{google2017}, the federated learning system was introduced to address the issue of data privacy. Therein, the mobile devices perform computation of model training locally on their training data according to the model released by the model owner. Such design enables mobile device to collaboratively learn a shared prediction model while keeping all the training data on the device~\cite{algorithmia2018}. However, the independent and rational mobile devices need incentive to participate in federated learning. In practice, asking the mobile devices to work as sacrificial volunteers is not an economically viable and sustainable option. Moreover, in the federated learning paradigm, the direct communication between model owner and mobile devices is still required for transferring model updates, i,e., Fig.~\ref{fig:system_model}(a). In many scenarios, the direct communication may be unavailable because of limited transmission range and energy inefficiency because of high transmission power.


To address the incentive issues of the mobile devices, we adopt the service pricing scheme to motivate the mobile devices to participate in federated learning. Under the service pricing scheme, the machine learning is provided by the mobile devices as a service to the model owner. Then, the learning service, i.e., model update generation and trading as well as data collection, is performed in a decentralized manner. Additionally, to overcome the energy inefficiency in the model update transfer for the mobile devices, we resort to relay networking to ensure that the model updates are transferred in a cooperative manner. The mobile devices cooperatively form a relay network by providing relay service to each other and directly or indirectly connect to the access point of the model owner for model update transfer, e.g., Fig.~\ref{fig:system_model}(b). Note here that the mobile device that acts as the relay node will use the average operator to combine its own model update with its received model updates. Then, the file size of the model update is not affected, and hence providing relay service does not significantly affect the energy consumption of the mobile devices.


In this paper, we propose a novel framework of cooperative federated learning system. Our designed federated learning system involves two parties, i.e., the massive-scale mobile devices working as learning service providers, and the model owner handling the learning task dispatching (model releasing) and model updates collection. The mobile devices price their learning service by deciding on the price of one unit of their training data. In return, the model owner determines the size of training data for each mobile device. As such, under the service pricing scheme, the mobile devices optimize the prices of their data to motivate the model owner to determine a larger size of training data and hence maximize their profits. However, under the cooperative relay network design, the larger size of training data implies the lower probability of enjoying the relay service. As a result, the learning service pricing and cooperative relaying should be considered jointly. By using the pricing-based data rent and a self-organized relay network design shown in Fig~\ref{fig:system_model}(b) for federated learning, the following key properties are guaranteed in our system:
\begin{itemize}

\item The model update throughput of the model owner scales well such that the massive model update volume from the mobile devices is handled smoothly.

\item The rational and self-interested mobile devices noncooperatively decide on their own price of one unit of training data for individual profit optimization and cooperatively transfer their model updates.

\item Significantly reduces the congestion in the communication for both the model owner and the mobile devices.

\end{itemize}

The rest of the paper is organized as follows. Section~\ref{sec:model} describes the system model. Section~\ref{sec:game_formulation} presents the formulation of a Stackelberg game. Section~\ref{sec:equilibrium_analysis} analyzes the equilibrium of the proposed Stackelberg game. Section~\ref{sec:performance} presents the numerical performance evaluation. Section~\ref{sec:conclusion} concludes the paper.

\section{System Description}
\label{sec:model}

\begin{figure}[!]
 \centering
 \includegraphics[width=0.45\textwidth,trim=150 160 225 150,clip]{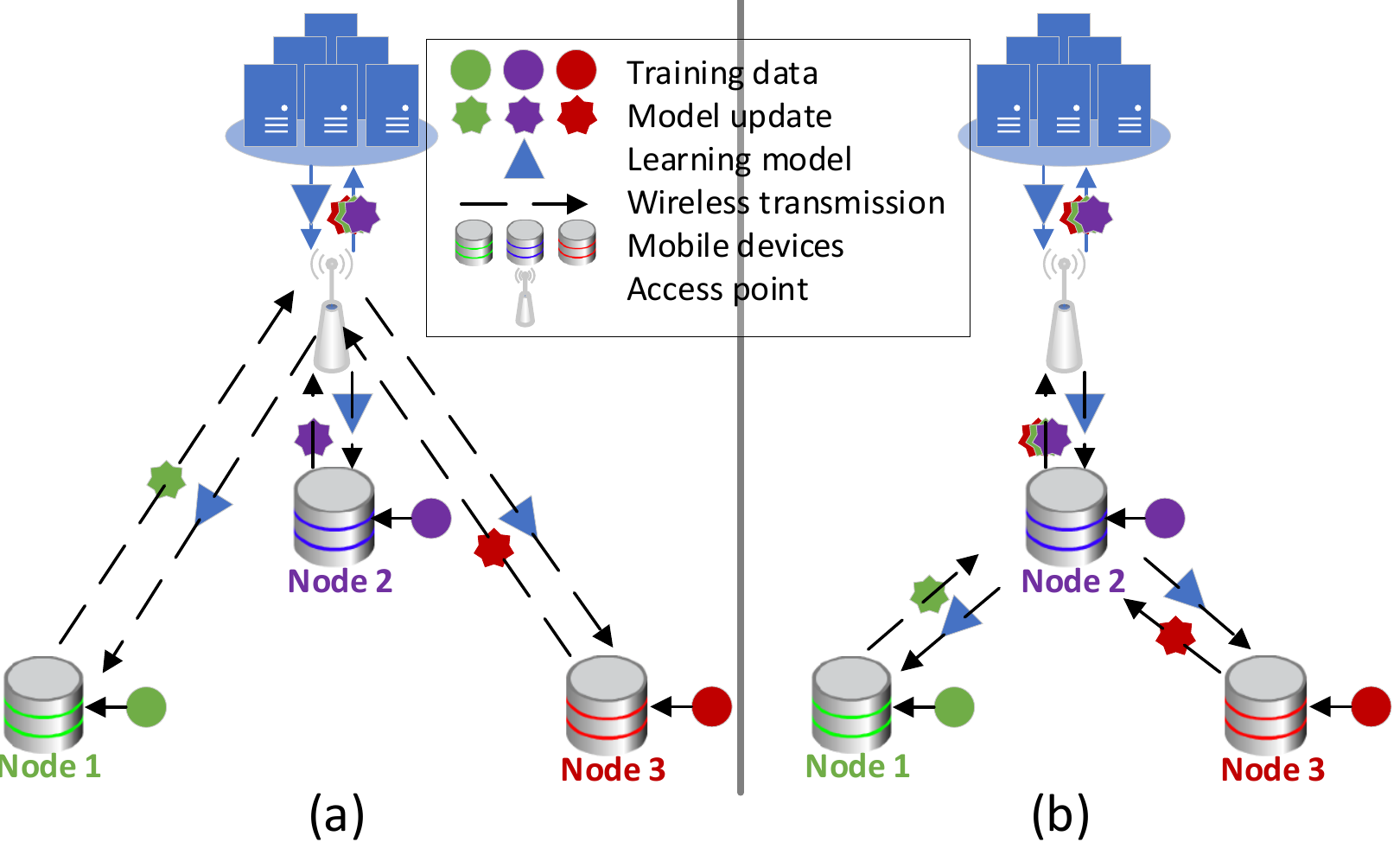}
 \caption{(a) Original communication scenario for federated learning; (b) Our designed communication scenario for federated learning.}
 \label{fig:system_model}
\end{figure}

\begin{table}[!]\label{definition}
\centering
\caption{Notations}
\begin{tabular}{|c|l|}
\hline
\hline
{\bf{Symbol}} & {\bf{Description}} \\
\hline
$i$, $N_{\rm{D}}$	&	Node $i$ and the access point of the model owner.	\\
\hline
${\cal{N}}^{\rm{0}}$	&	The set of mobile devices.\\
\hline
${\cal{N}}$, $\left|{\cal{N}}\right|$	&	\makecell[l]{${\cal{N}}={\cal{N}}^{\rm{0}}  \cup\left\{N_{\rm{D}}\right\}= \left\{1,2,\ldots, \left|{\cal{N}}^{\rm{0}}\right|, N_{\rm{D}}\right\}$ is a sorted set\\ and its cardinality.}\\
\hline
$P_{ij}$	&	\makecell[l]{The transmission power used by mobile device $i$ to transfer \\the model update to mobile device $j$.}	\\
\hline
$T^{\rm{s}}_i$, $T^{\rm{a}}_i$	&	\makecell[l]{The time used by mobile device $i$ for processing the model \\update and average operator, respectively.}\\
\hline
$h_{ij}$, $d_{ij}$ & \makecell[l]{The channel gain of the transmission between mobile \\devices $i$ and $j$, and the distance between mobile \\devices $i$ and $j$.} \\
\hline
$q_i$, $s^{\rm{d}}_i$ & \makecell[l]{The price of one unit of training data for mobile device $i$ \\and the size of the data set used by mobile device $i$ for \\generated the model update. }\\
\hline
$c^{\rm{p}}_i$, $c^{\rm{t}}_i$	&	\makecell[l]{The costs that mobile device $i$ uses one unit of energy for \\processing one unit of data and transferring, respectively.} \\
\hline
$c^{\rm{a}}$ & \makecell[l]{The fee that each mobile device $i \in {\cal{N}}$ charged by its rely \\node for the relay service.}\\
\hline
$I^{\rm{d}}$ & The size of the model update.\\
\hline
$r_i$ & The transmission rate of mobile device $i$.\\
\hline
\end{tabular}
\end{table}

We consider a cooperative federated learning system as shown in Fig.~\ref{fig:system_model}(b). Specifically, a model owner employs a set of mobile devices, e.g., mobile phones, to train a high-quality centralized model~\cite{konecny2016federated}. The set of mobile devices is denoted by ${\cal{N}}^{\rm{0}}$, e.g., mobile devices $1$, $2$, and $3$ in Fig.~\ref{fig:system_model}(b). Each mobile device $i \in {\cal{N}}^{\rm{0}}$ uses a part of its data and performs computation on its data locally to generate the model update for training the model of the model owner. The model owner negotiates with the mobile devices about the size of their training data, i.e., $s^{\rm{d}}_i$. In return, each mobile device $i \in {\cal{N}}^{\rm{0}}$ will receive the revenue $q_is^{\rm{d}}_i$ from the model owner, where $q_i$ is the price for one unit of mobile device $i$'s training data. Intuitively, the learning accuracy of the model depends on the total size of all the mobile devices' training data. Specifically, the learning accuracy of the model becomes higher as the total size of all the mobile devices' training data increases. In this case, we incorporate the results in~\cite{gu2001modelling} to describe the relationship between the learning accuracy of the model and the total size of all the mobile devices' training data. As a result, the utility of the model owner is defined as follows:
\begin{equation}\label{eq:utility_server}
\footnotesize{U\left({\bf{s}}^{\rm{d}}, {\bf{q}}\right)= f\left({\bf{s}}^{\rm{d}}\right) - \sum\limits_{i\in{\cal{N}}^{\rm{0}}}q_i s^{\rm{d}}_i,}
\end{equation}
where $f\left(\cdot\right)$ is the function describing the relationship between the learning accuracy of the model and the total size of all the mobile devices' training data~\cite{gu2001modelling}. Note here that $f\left(\cdot\right)$ is an increasing concave function implying that the learning accuracy of the model keeps increasing as the size of training data increases while the marginal increasing speed of the learning accuracy of the model decreases~\cite{gu2001modelling}.

The mobile devices can cooperate with each other for transferring their model updates to the access point of the model owner. Let $P_{ij}$ denote the transmission power used by mobile device $i$ for transferring its model update to mobile device $j$, where we define the transmission power matrix ${\bf{P}}=\left[p_{ij}\right]_{i,j\in{\cal{N}}}$. $I_{ij}$ is the element of the indicator matrix ${\bf{I}}$, i.e., ${\bf{I}}=\left[I_{ij}\right]_{i,j\in{\cal{N}}}$, and defined as follows:
\begin{equation}
\footnotesize{I_{ij}=\left\{\begin{aligned}
{1},&\,{P_{ij}>0}&\\
{0},&\,{\text{Otherwise}}&
\end{aligned}\right.,\, \forall i,\, j \in {\cal{N}}}.
\end{equation}
To provide the learning service, mobile device $i$ has the cost of ${c_i^{\rm{p}}s^{\rm{d}}_i}$ due to the energy consumption incurred from the computation. Moreover, each mobile device $i\in{\cal{N}}^{\rm{0}}$ has another cost incurred by the wireless transmission of its model update. Let $c^{\rm{t}}_i$ be the cost that mobile device $i$ uses one unit of power for wireless transmission, the instantaneous cost for mobile device $i$ incurred by the transmission is ${c_i^{\rm{t}}\sum\limits_{j \in {\cal{N}}} {{P_{ij}}} }$. Under the assumption that the sizes of the model updates transferred by the mobile devices are the same, the energy consumption from the wireless transmission depends on the transmission rate of each mobile device, i.e., $r_i$. We denote the size of the model update by $I^{\rm{d}}$. The time for transferring the model update is accordingly $\frac{I^{\rm{d}}}{r_i}$, and hence the energy consumption of mobile device $i$ incurred by the wireless transmission is ${c_i^{\rm{t}}\frac{I^{\rm{d}}}{r_i}\sum\limits_{j \in {\cal{N}}} {{P_{ij}}} }$. Furthermore, with the relay service among the mobile devices, each mobile device $i \in {\cal{N}}^{\rm{0}}$ will have a revenue of ${{c^{\rm{a}}}\sum\limits_{j \in {\cal{N}}^{\rm{0}}} {{I_{ji}}} }$ due to its provided relay service while a cost of ${{c^{\rm{a}}}\left(1-I_{iN_{\rm{D}}}\right) }$ incurred by the use of the relay service. ${{c^{\rm{a}}}\left(1-I_{iN_{\rm{D}}}\right) }$ comes from the fact that mobile device $i$ uses the relay service and hence pay the fee ${c^{\rm{a}}}$ if it does not directly transfer the model update to the model owner, i.e., $I_{iN_{\rm{D}}}=0$. Finally, with the revenue of providing the learning service to the model owner, i.e., ${q_i}s_i^{\rm{d}}$, the corresponding profit of mobile device $i$ is
\begin{equation}\label{eq:mobile device_profit}
\footnotesize{\begin{aligned}
&{\Pi _i} \left({\bf{P}}, q_i, s^{\rm{d}}_i\right)= {q_i}s_i^{\rm{d}} -  {c_i^{\rm{t}}\frac{I^{\rm{d}}}{r_i}\sum\limits_{j \in {\cal{N}}} {{P_{ij}}} } -  {c_i^{\rm{p}}s^{\rm{d}}_i}\\
&+ {{c^{\rm{a}}}\sum\limits_{j \in {\cal{N}}^{\rm{0}}} {{I_{ji}}} } - {{c^{\rm{a}}}\left(1-I_{iN_{\rm{D}}}\right) }.
\end{aligned}}
\end{equation}

To form the wireless relay networks among the mobile devices and hence achieve the energy efficient communication for the model update transfer, we have two sets of constraints. The first set of the constraints is for the routing. The second set of the constraints is for the model update arrival time at the relay point.

For the routing, we first have the constraint to ensure that every mobile device can connect to and transfer its model update to only one of other mobile devices or directly to the access point of the model owner. The constraint is expressed as follows:
\begin{equation}\label{con:routing_1}
\footnotesize{\sum\limits_{j \in {\cal{N}}} {{I_{ij}}}  = 1 \quad {\text{and}}\quad I_{ii}=0, \, \forall i \in {\cal{N}}^{\rm{0}}.}
\end{equation}

Secondly, we have the constraint that at least one mobile device connects to the access point of the model owner and acts as one of the last-hop nodes. Otherwise, none of the mobile devices can transfer the model update to the model owner. The corresponding constraint is expressed as follows:
\begin{equation}\label{con:routing_2}
\footnotesize{\sum\limits_{i\in{\cal{N}}^{\rm{0}}}I_{iN_{\rm{D}}} \ge 1.}
\end{equation}

We then have the constraint that the model update of each mobile device can finally and affirmatively arrive at the model owner. That is, each mobile device can transfer its model update to the model owner after a limited number of mobile devices as its relay nodes, i.e.,
\begin{equation}\label{con:routing_3}
\footnotesize{{\bf{I}}^{\left|{\cal{N}}^{\rm{0}}\right|-1}={\bf{I}}_{\rm{D}},}
\end{equation}
where
\begin{equation}
\footnotesize{{{\bf{I}}_{\rm{D}}}^\top = \left[
\begin{aligned}
\quad{0}&\quad{\cdots}&\,{0}\quad\\
\quad{\vdots}&\quad{\ddots}&\,{\vdots}\quad\\
\quad{0}&\quad{\cdots}&\,{0}\quad\\
\quad{1}&\quad{\cdots}&\,{1}\quad
\end{aligned}
\right]
\begin{aligned}
\leftarrow&1\\&\vdots \\ \leftarrow&\left|{\cal{N}}^{\rm{0}}\right|\\ \leftarrow&N_{\rm{D}}
\end{aligned}.}
\end{equation}

Regarding the constraint for the model update arrival time on the relay node, we have the constraint in~(\ref{con:arriving_time}). This constraint is to ensure that the model update of mobile device $i$ will arrive at mobile device $j$ before mobile device $j$ finishes the computing of its model update if mobile device $i$ wants to transfer its model update by choosing mobile device $j$ as its relay node. The time used by mobile device $i$ for providing the learning service can be divided into three periods. The first time period, denoted by $T^{\rm{s}}_i$, is for performing the computation on the training data according to the model. The model update is generated at the end of this time period. Suppose that $r^{\rm{p}}_i$ is the processing rate of mobile device $i$, $T^{\rm{s}}_i$ is accordingly defined as $T^{\rm{s}}_i = \frac{s^{\rm{d}}_i}{r^{\rm{p}}_i}$~\cite{gharaibeh2013efficient}. The second time period is for mobile device $i$ to provide relay service to other mobile devices. For each received model update from another mobile device, mobile device $i$ needs to spend the time of $T^{\rm{a}}_i$ for combining the received model update with its own model update by using the average operator. The length of the second time period is therefore linearly related to the number of model updates received by mobile device $i$, i.e., the number of mobile devices that use mobile device $i$'s relay service $\sum\limits_{j\in {\cal{N}}}I_{ji}$, and expressed as $T^{\rm{a}}_i \sum\limits_{j\in {\cal{N}}}I_{ji}$. The last time period is for mobile device $i$ to transfer its model update and expressed as $\frac{I^{\rm{d}}}{r_i}$, where $r_i$ is the transmission rate of mobile device $i$ defined in~(\ref{eq:transmission_rate_i}). To ensure that mobile device $i$ can successfully use the relay service from other mobile devices, the sum of three time periods at mobile device $i$ must be shorter than the first time period of its relay node, i.e.,
\begin{equation}\label{con:arriving_time}
\footnotesize{T^{\rm{s}}_i+T^{\rm{a}}_i \sum\limits_{j\in {\cal{N}}^{\rm{0}}}I_{ji}+\frac{I^{\rm{d}}}{r_i}  \le \sum\limits_{j \in {\cal{N}}^{\rm{0}}} I_{ij} T^{\rm{s}}_j,\, \forall i \in {\cal{N}}^{\rm{0}}}.
\end{equation}

We assume that the mobile devices using the same relay node share the channel and hence generate mutual interference. Let $h_{ij}$ denote the channel gain from mobile device $i$ to mobile device $j$, $d_{ij}$ be the distance between mobile devices $i$ and $j$, and $\alpha \ge 2$ be the path loss coefficient of wireless communication. Then, we define a matrix ${\bf{H}}=\left[H_{ij}\right]_{i,j\in{\cal{N}}}$, the element of which, i.e., $H_{ij}$, is $\frac{h_{ij}}{\left(d_{ij}\right)^\alpha}$. Accordingly, the propagation model for data transmission of mobile device $i$, i.e., the transmission rate of mobile device $i$, $r_i$ is given by
\begin{equation}\label{eq:transmission_rate_i}
\footnotesize{\begin{aligned}
&{r_i} = {w_i}{\log _2}\left( 1+{\frac{{\sum\limits_{j\in{\cal{N}}} {\frac{{{h_{ij}}}}{{{{\left( {{d_{ij}}} \right)}^\alpha }}}{P_{ij}}} }}{{{{\left( {{{\bf{H}}_{\bf{I}}}\left( {:,i} \right)} \right)}^\top}\left( {{{\bf{P}}_{\bf{I}}}\left( {:,i} \right)} \right) - \sum\limits_{j\in{\cal{N}}} {\frac{{{h_{ij}}}}{{{{\left( {{d_{ij}}} \right)}^\alpha }}}{P_{ij}}}  + {\sigma ^2}}}} \right)\\
&= {w_i}{\log _2}\left( 1+{\frac{{  \left({\bf{H}}\left(i,:\right)\right)^\top \left({\bf{P}}\left(i,:\right) \right)}}{{{{\left( {{{\bf{H}}_{\bf{I}}}\left( {:,i} \right)} \right)}^\top}\left( {{{\bf{P}}_{\bf{I}}}\left( {:,i} \right)} \right) - { \left({\bf{H}}\left(i,:\right)\right)^\top \left({\bf{P}}\left(i,:\right)\right)}  + {\sigma ^2}}}} \right),
\end{aligned}}
\end{equation}
where $\sigma^2$ is the noise, $w_i$ is the bandwidth of mobile device $i$, ${{\bf{P}}_{\bf{I}}} = {{\bf{P}}{\bf{I}}^\top}$, and ${{\bf{H}}_{\bf{I}}} = {{\bf{H}}{\bf{I}}^\top}$. Note here that we use ${\bf{H}}\left(i,:\right)$ and ${\bf{H}}\left(:,i\right)$ to represent the $i$-th row vector and $i$-th column vector of ${\bf{H}}$, respectively, and the usage of this notation is defined similarly in the rest of this paper.

\section{Stackelberg game Formulation}
\label{sec:game_formulation}

In this section, we model the interaction between the mobile devices and the model owner as a Stackelberg game. In the Stackelberg game, the model owner is the buyer as it uses the learning service provided by the mobile devices. Then, the mobile devices that are the service providers act as the sellers. The sellers typically make their decisions before the buyers. Following this case, the model owner inherently acts as the single follower in the lower level of the Stackelberg game while the mobile devices are the corresponding leaders. In the lower level of the game, i.e., the lower-level subgame, the model owner determines the size of training data for the mobile devices. In the upper level, i.e., the upper-level subgame, the mobile devices decide on the price for one unit of their training data. Moreover, since the mobile devices cooperatively send their model update to the model owner, each mobile device also needs to independently decide on its relay node as well as its transmission power. As a result, the Stackelberg game can be formally defined as follows:
\begin{enumerate}

\item \textsl{Lower-level subgame}: Given the fixed vector of the prices of one unit of training data ${\footnotesize{{\bf{q}}=\left[q_1, \ldots,q_{\left|{ \cal{N}}^{\rm{0}}\right|}\right]^\top}}$, the lower-level subgame is defined by the a three-tuple ${\cal{G}}_{\rm{L}}=\left\{{\bf{s}}^{\rm{d}}, {\mathbb{S}}^{\rm{d}}, {U}\right\}$, where
    \begin{itemize}

    \item ${\footnotesize{{\bf{s}}^{\rm{d}}=\left[s^{\rm{d}}_1,\ldots,s^{\rm{d}}_{\left| { \cal{N}}^{\rm{0}} \right|}\right]^\top}}$ is the vector of the sizes of training data;

    \item ${\mathbb{S}}^{\rm{d}} = \left\{\left[s^{\rm{d}}_1,\ldots,s^{\rm{d}}_{\left| { \cal{N}}^{\rm{0}} \right|}\right]^\top\left|s^{\rm{d}}_i \in \left[0, s^{{\rm{d}},{\rm{u}}}_i\right], i \in {\cal{N}}^{\rm{0}}\right.\right\} \\ \subset {\mathbb{R}}^{\left| { \cal{N}}^{\rm{0}} \right|}$ is the domain of definition for ${\bf{s}}^{\rm{d}}$ and an M-polyhedron, where $s^{{\rm{d}},{\rm{u}}}_i$ is the upper bound of $s^{{\rm{d}}}_i$;

    \item ${U} = U\left({\bf{s}}^{\rm{d}}, {\bf{q}}\right)$ is the utility function of the model owner defined in~(\ref{eq:utility_server});

    \end{itemize}

\item \textsl{Upper-level subgame}: After the model owner's demand of data size ${\bf{s}}^{\rm{d}}$ is determined in the lower-level subgame, the mobile devices form a upper-level subgame defined by a six-tuple ${\cal{G}}_{\rm{U}}=\left\{{\cal{N}}^{\rm{0}}, {\bf{P}}, {\mathbb{P}}, {\bf{q}}, {\mathbb{Q}}, {\bf{\Pi}}\right\}$, where
    \begin{itemize}

    \item ${\cal{N}}^{\rm{0}}$ is the set of the mobile devices;

    \item ${\bf{P}}=\left[P_{ij}\right]_{i,j\in{\cal{N}}}$ is the matrix of the power for wireless transmission;

    \item ${\mathbb{P}}=\left\{\left[P_{ij}\right]_{i,j \in {\cal{N}}}\left|P_{ij}\in\left[0,P^{\rm{u}}_{ij}\right], i,j \in {\cal{N}}\right.\right\} \subset {\mathbb{R}}^{\left| { \cal{N}} \right|\times \left| { \cal{N}} \right|}$ is the domain of definition for ${\bf{P}}$, where $P^{\rm{u}}_{ij}$ is the upper bound of $P_{ij}$;

    \item ${\bf{q}}=\left[q_1, \ldots,q_{\left|{ \cal{N}}^{\rm{0}}\right|} \right]^\top$ is the vector of prices for one unit of training data.

    \item ${\mathbb{Q}} = \left\{\left[q_1, \ldots,q_{\left|{ \cal{N}}^{\rm{0}}\right|} \right]^\top\left|q_i \in \left[0, q^{{\rm{u}}}_i\right], i \in {\cal{N}}^{\rm{0}}\right.\right\}\subset {\mathbb{R}}^{\left| { \cal{N}}^{\rm{0}} \right|}$ is the domain of definition for ${\bf{q}}$ and an M-polyhedron, where $q^{{\rm{u}}}_i$ is the upper bound of $q_i$;

    \item  ${\bf{\Pi}}=\left[\Pi_1,\ldots,\Pi_{\left|{\cal{N}}^{\rm{0}} \right|}\right]^\top$ is the vector of the profits for the mobile devices, where $\Pi_i={\Pi _i} \left({\bf{P}}, q_i, s^{\rm{d}}_i\right)$ is the profit of mobile device $i$ defined in~(\ref{eq:mobile device_profit}).

    \end{itemize}

\end{enumerate}

Based on the game formulation, we consider a Stackelberg equilibrium to be the solution for the model owner and the mobile devices.

\section{Equilibrium Analysis}
\label{sec:equilibrium_analysis}

By following the backward induction, we firstly use the first-order optimality condition to obtain the optimal solution to the lower-level subgame ${\cal{G}}_{\rm{L}}$. The existence of the optimal solution to the lower-level subgame ${\cal{G}}_{\rm{L}}$ is proven to exist by showing the concavity of its utility function. This optimal solution is further proven to be unique by showing the negative definiteness of the Hessian matrix of the utility function of the lower-level subgame ${\cal{G}}_{\rm{L}}$. Then, we substitute the NE of the lower-level subgame ${\cal{G}}_{\rm{L}}$ into the upper-level subgame ${\cal{G}}_{\rm{U}}$ and investigate the solution to the upper-level subgame ${\cal{G}}_{\rm{U}}$ by capitalizing on the exterior point method.

\subsection{Solution to Lower-level Subgame}

To find an optimal solution for the lower-level subgame $G_{\rm{L}}$, we need to take the first derivative of the utility function of the model owner given in~(\ref{eq:utility_server}) with respect to $s_i^{\rm{d}}$ as follows:
\begin{equation}
\footnotesize{\begin{aligned}
&\frac{\partial }{{\partial s_i^{\rm{d}}}}U\left( {{{\bf{s}}^{\rm{d}}},{\bf{q}}} \right) = \frac{\partial }{{\partial s_i^{\rm{d}}}}\left[ {  {{f}} \left( {{\bf{s}}^{\rm{d}}} \right) - \sum\limits_{i \in {{\cal{N}}^{\rm{0}}}} {{q_i}} s_i^{\rm{d}}} \right]\\
=&  - {q_i} + \frac{\partial }{{\partial s_i^{\rm{d}}}}{f}\left( {{\bf{s}}^{\rm{d}}} \right), \,\forall i \in {\cal{N}}^{\rm{0}},
\end{aligned}}
\end{equation}
where $f\left({\bf{s}}^{\rm{d}}\right)=\sum\limits_{i\in{\cal{N}}^{\rm{0}}} f_i \left( s_i^{\rm{d}}\right)$. Without loss of generality and for the trackable analysis, we adopt Weibull model as suggested in~\cite{gu2001modelling}, i.e., ${f_i}\left( {s_i^{\rm{d}}} \right)=a_i-b_i \exp\left(-c_i {s_i^{\rm{d}}} \right)$. Let $\frac{\partial }{{\partial s_i^{\rm{d}}}}U\left( {{{\bf{s}}^{\rm{d}}},{\bf{q}}} \right)=0$, $\forall i \in {\cal{N}}^{\rm{0}}$, we have ${s_i^{\rm{d}}}^*$, i.e.,  the best response of the model owner, as follows:
\begin{equation}\label{eq:optimal_demand}
\footnotesize{\begin{aligned}
 &- {q_i} + \frac{\partial }{{\partial s_i^{\rm{d}}}}{f}\left( {{\bf{s}}^{\rm{d}}} \right) =  - {q_i} + \frac{\partial }{{\partial s_i^{\rm{d}}}}\left[ {{a_i} - {b_i}\exp \left( { - {c_i}{s_i^{\rm{d}}}} \right)} \right]\\
 =&  - {q_i} + {c_i}{b_i}\exp \left( { - {c_i}{s_i^{\rm{d}}}} \right) = 0
 \,\Leftrightarrow \,{s_i^{\rm{d}}}^* = \frac{1}{{{c_i}}}\ln \frac{{{c_i}{b_i}}}{{{q_i}}}
 \end{aligned}}
\end{equation}
Due to the strict concavity of $f_i\left(\cdot\right)$ and the linearity of $-q_is^{\rm{d}}_i$ with respect to $s^{\rm{d}}_i$, $U\left( {{{\bf{s}}^{\rm{d}}},{\bf{q}}} \right)$ is therefore concave with respect to $ s_i^{\rm{d}}$ and its concavity indicates the existence of the solution to the lower-level subgame ${\cal{G}}_{\rm{L}}$. Moreover, with negative elements on the primary diagonal and zero elements on the off-diagonal, the Hessian matrix of $U\left( {{{\bf{s}}^{\rm{d}}},{\bf{q}}} \right)$ is negative definite, and hence the solution to the lower-level subgame ${\cal{G}}_{\rm{L}}$ is unique.

\subsection{Solutions to Upper-level Subgame}

After obtaining the optimal demand of the data size for the model owner, we investigate the upper-level subgame ${\cal{G}}_{\rm{U}}$ for the mobile devices. At the Nash Equilibrium (NE), no player can increase its profit by choosing a different strategy provided that the other players' strategy is unchanged~\cite{osborne2004introduction}. We firstly substitute the optimal demand of the data size for the model owner given in~(\ref{eq:optimal_demand}) into the profit functions of the mobile devices given in~(\ref{eq:mobile device_profit}) and have the new profit functions for the mobile devices as follows:
\begin{equation}\label{eq:profit_i_constrained}
\footnotesize{\begin{aligned}
{\Pi _i}\left( {\bf{P}},{q_i} \right) =& {\Pi _i}\left( {{\bf{P}},{q_i},{s_i^{\rm{d}}}^*} \right) = {q_i}\frac{1}{{{c_i}}}\ln \frac{{{c_i}{b_i}}}{{{q_i}}} - c_i^{\rm{t}}\frac{{{I^{\rm{d}}}}}{{{r_i}}}\sum\limits_{j \in {\cal{N}}} {{P_{ij}}}  \\
&- c_i^{\rm{p}}\frac{1}{{{c_i}}}\ln \frac{{{c_i}{b_i}}}{{{q_i}}} + {c^{\rm{a}}}\sum\limits_{j \in {{\cal{N}}^{\rm{0}}}} {{I_{ji}}}  - {c^{\rm{a}}}\left(1-I_{i N_{\rm{D}}}\right)
\end{aligned}}
\end{equation}
with the constraints~(\ref{con:routing_1})-(\ref{con:arriving_time}). As the constraints~(\ref{con:routing_1})-(\ref{con:arriving_time}) are nonlinear, we adopt the exterior point method~\cite{yang1994exterior}. Then, we can rewritten the constrained profit function of mobile device $i\in{\cal{N}}^{\rm{0}}$, i.e.,~(\ref{eq:profit_i_constrained}), into a unconstrained objective function as follows:
\begin{equation}
\footnotesize{\begin{aligned}
&{\Pi _i}\left( {{\bf{P}},{q_i}}, M\right)
= {q_i}\frac{1}{{{c_i}}}\ln \frac{{{c_i}{b_i}}}{{{q_i}}} - c_i^{\rm{t}}\frac{{{I^{\rm{d}}}}}{{{r_i}}}\sum\limits_{j \in {\cal{N}}} {{P_{ij}}}  - c_i^{\rm{p}}\frac{1}{{{c_i}}}\ln \frac{{{c_i}{b_i}}}{{{q_i}}} \\
&+ {c^{\rm{a}}}\sum\limits_{j \in {{\cal{N}}^{\rm{0}}}} {{I_{ji}}}  - {c^{\rm{a}}}\left( {1 - {I_{i{N_{\rm{D}}}}}} \right) +M \rho\left( {{\bf{P}},{q_i}}\right),
\end{aligned}}
\end{equation}
where $M$ is the penalty coefficient with huge positive value and
\begin{equation}\label{eq:profit_i_exterior_point}
\footnotesize{\begin{aligned}
&\rho\left( {{\bf{P}},{q_i}}\right)=\left[- {\left( {\sum\limits_{j \in {\cal{N}}} {{I_{ij}}}  - 1} \right)^2} - {\left( {{I_{ii}}} \right)^2} - {\left( {{{\bf{I}}^{\left| {\cal{N}} \right| - 1}} - {{\bf{I}}_{\rm{D}}}} \right)^2}\right.\\
&\left.+ \left( {\sum\limits_{i \in {\cal{N}}} {{I_{i{N_{\rm{D}}}}}}  - 1} \right) + \left( {\sum\limits_{j \in {\cal{N}}} {{I_{ij}}} T_j^{\rm{s}} - T_i^{\rm{s}} - T_i^{\rm{a}}\sum\limits_{j \in {\cal{N}}} {{I_{ji}}}  - \frac{{{I^{\rm{d}}}}}{{{r_i}}}} \right)\right].
\end{aligned}}
\end{equation}


\section{Performance Evaluation}
\label{sec:performance}

In this section, we present numerical studies to evaluate the performance of the cooperative federated learning system. For the ease of illustration, we consider $9$ mobile devices, i.e, $\left|{\cal{N}}^{\rm{0}}\right| = 9$, working as the learning entities. The bandwidth and channel gain are respectively $h_{ij} = 10$ and $w_i=1$, $\forall i, \, j\in {\cal{ N}}^{\rm{0}}$, and the noise $\sigma^2$ is $1$. The size of the model update is assumed to be $I^{\rm{d}}=0.1$. The distances among the mobile devices as well as that between the mobile devices and the access point of the model owner, i.e., $d = \left[d_{ij}\right]_{i,j\in {\cal{N}}}$ , follows a uniform distribution under the plane of $\left[0,10\right] \times \left[0,10\right]$. The vector of the costs of consuming one unit of power for wireless communication is generated by using Gaussian distribution and shown as follows: ${\bf{c}}^{\rm{t}} = \left[c^{\rm{t}}_i\right]_{i\in{\cal{N}}^{\rm{0}}} = \left[58, 61, 51.5, 58.5, 95, 46, 175, 124.5, 31\right]^\top$. Similarly, the following parameters are generated by using Gaussian distribution. The vector of the costs of processing one unit of data is ${\bf{c}}^{\rm{p}} = \left[c^{\rm{p}}_i\right]_{i\in{\cal{N}}^{\rm{0}}} =\left[4.3, 8.5, 13.6, 9.5, 9.8, 6.7, 8.1, 5.5, 11.2\right]\times10^{-3}$. We set the fee of relay service to be $c^{\rm{a}}=0.0096$. The vector of processing rate is ${\bf{r}}^{\rm{p}} = \left[r^{\rm{p}}_i\right]_{i\in{\cal{N}}^{\rm{0}}}=\left[8.81, 8.93, 9.725, 6.165, 4.15, 4.195, 6.525, 8.215, 5.105, \right.\\\left. 5.96\right]^\top\times10$. The vector of the time of using average operator for the mobile devices is ${\bf{T}}^{\rm{a}} = \left[T^{\rm{a}}_i\right]_{i\in{\cal{N}}^{\rm{0}}}=\left[1.21, 1.29, 0.53, 1.07, 1.07, 0.95, 1.3, 0.88, 0.72\right]\times10^{-2}$. The coefficient of $f_i\left(\cdot\right)$ defined in~(\ref{eq:utility_server}) is ${\bf{c}}=\left[c_{ij}\right]_{i,j\in{\cal{N}}^{\rm{0}}}=\left[15.28, 9.17, 14.31, 11.21, 9.12, 13.61, 13.27, 9.63, 14.32\right]$ and
${\bf{a}}=\left[a_{ij}\right]_{i,j\in{\cal{N}}^{\rm{0}}}={\bf{b}}=\left[b_{ij}\right]_{i,j\in{\cal{N}}^{\rm{0}}}=\left[9.78, 9.15, 11.35, 11.17, 12.7, 9.15, 12.38, 13.5, 10.59\right]$.

\subsection{Numerical Result}

\begin{figure}[!]
 \centering
 \includegraphics[width=0.39\textwidth]{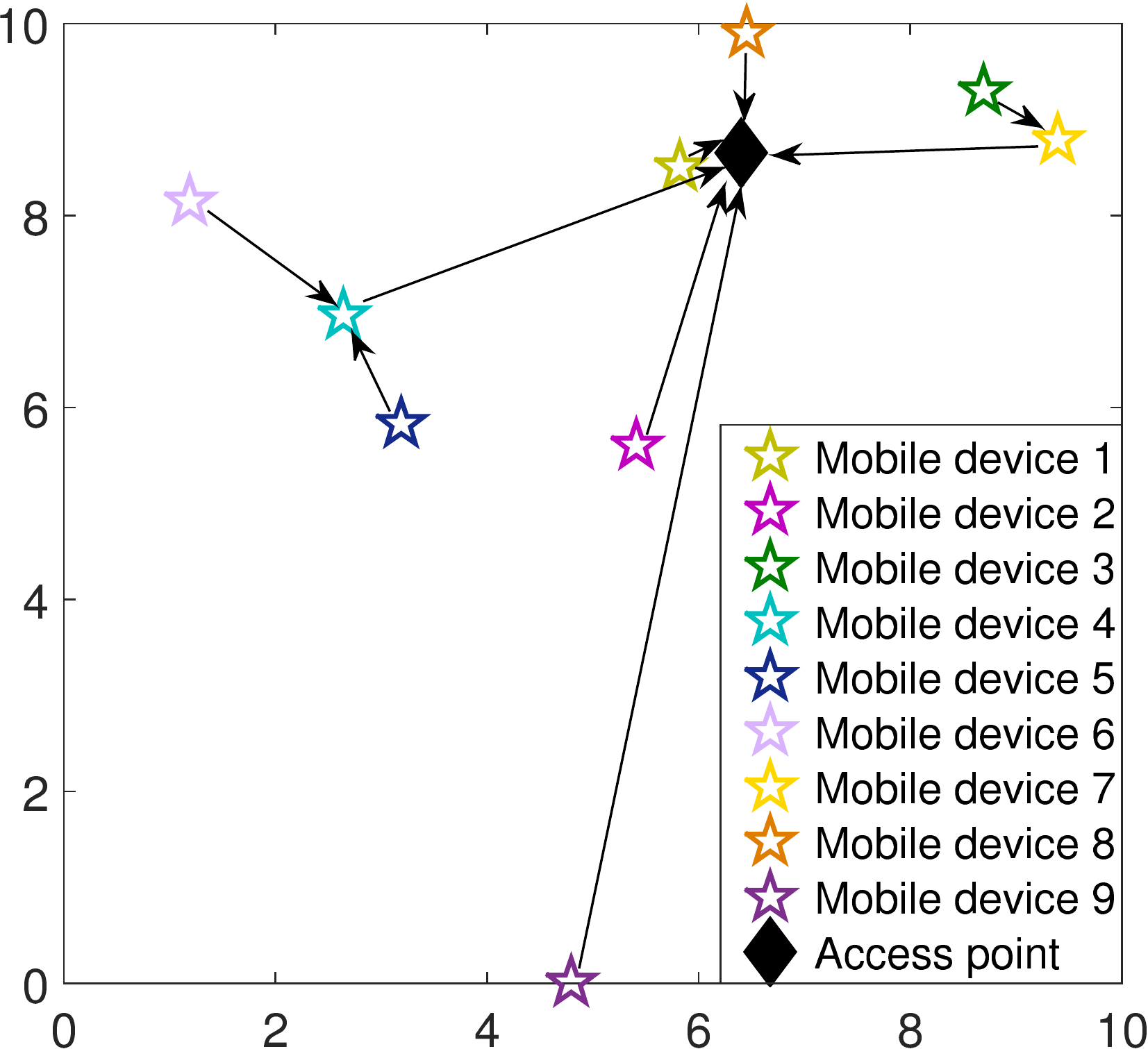}
 \caption{The wireless relay network for cooperative model update transfer.}
 \label{fig:connection}
\end{figure}

\begin{table}[!]
\centering
\caption{The routing result for cooperative model update transfer.}
\label{table:routing_result}
\begin{tabular}{|c|l|}
\hline
{\bf{Mobile device No.}} & {\bf{Routing}} \\
\hline
$1$	&	$1\rightarrow N_{\rm{D}}$.	\\
\hline
$2$	&	$2\rightarrow N_{\rm{D}}$.\\
\hline
$3$	&	$3\rightarrow 7 \rightarrow  N_{\rm{D}}$.\\
\hline
$4$	&	$4\rightarrow N_{\rm{D}}$.	\\
\hline
$5$	&	$5\rightarrow 4\rightarrow N_{\rm{D}}$.\\
\hline
$6$ & $6\rightarrow 4\rightarrow N_{\rm{D}}$. \\
\hline
$7$ & $7\rightarrow N_{\rm{D}}$.\\
\hline
$8$	&	$8\rightarrow N_{\rm{D}}$. \\
\hline
$9$ & $9\rightarrow N_{\rm{D}}$.\\
\hline
\end{tabular}
\end{table}

For the convenience to observe the routing result for the cooperative model update transfer, We present it in Fig.~\ref{fig:connection} and Table~\ref{table:routing_result}. As shown in Fig.~\ref{fig:connection}, the mobile devices self-organize the wireless rely network for cooperative model update transfer. For example, mobile device $3$ uses mobile device $7$ as the relay node for transferring the model update. This can significantly reduce the energy consumption of wireless communication for mobile device $3$. Moreover, both mobile devices $5$ and $6$ choose the same mobile device, i.e, mobile device $4$, as their relay node. This will incur the mutual interference between mobile devices $5$ and $6$ and hence result in energy inefficient communication. However, compared with choosing other mobile devices as the relay node or directly transferring the model update to the cental cloud, it is more energy efficient to choose mobile device $4$ as the relay node for mobile devices $5$ and $6$.

\begin{figure}[!]
 \centering
 \includegraphics[width=0.39\textwidth]{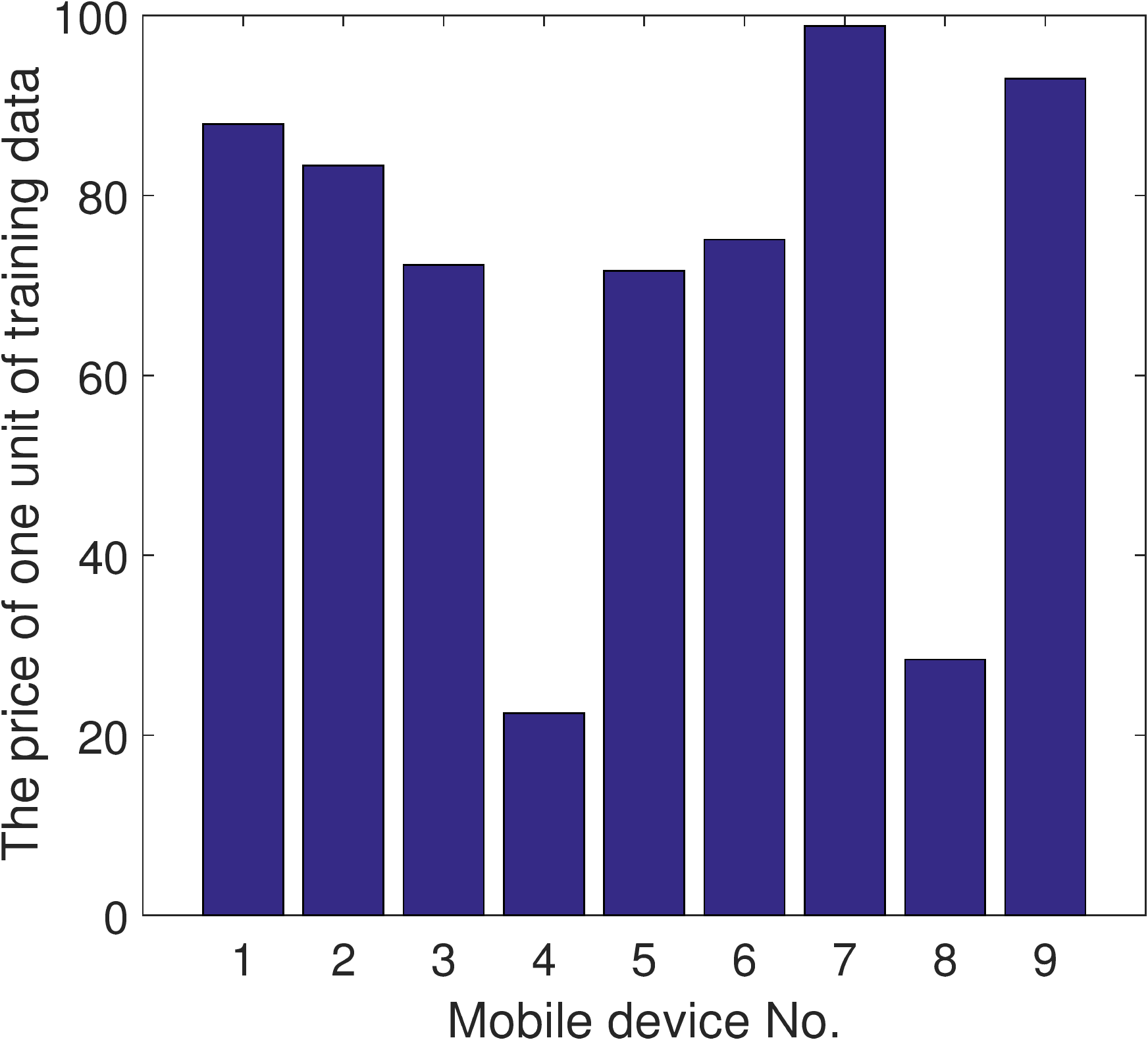}
 \caption{The price of one unit of the training data for each device.}
 \label{fig:price}
\end{figure}

We next evaluate the prices of one unit of training data for the mobile devices. As shown in Fig.~\ref{fig:price}, we observe that the price of one unit of training data for mobile device $7$ is the highest one among the mobile device. The reason is that the model update from mobile device $7$ received by the model owner is the combination of the model update from two mobile devices, i.e., mobile devices $7$ and $3$ as shown in Fig.~\ref{fig:connection} and Table~\ref{table:routing_result}. This means that the model update from mobile device $7$ contains much more valuable information than that generated by using the model update from only one mobile device, e.g., the model update of mobile devices $1$, $2$, and $8$. In contrast, although the model update from mobile device $4$ is the combination of the model updates from the model updates of mobile devices $4$, $5$, and $6$, the price of one unit of training data for mobile device $4$ is even less than that for mobile device $8$. This is due to the data quality and the preference of the model owner.

\begin{figure}[!]
 \centering
 \includegraphics[width=0.39\textwidth]{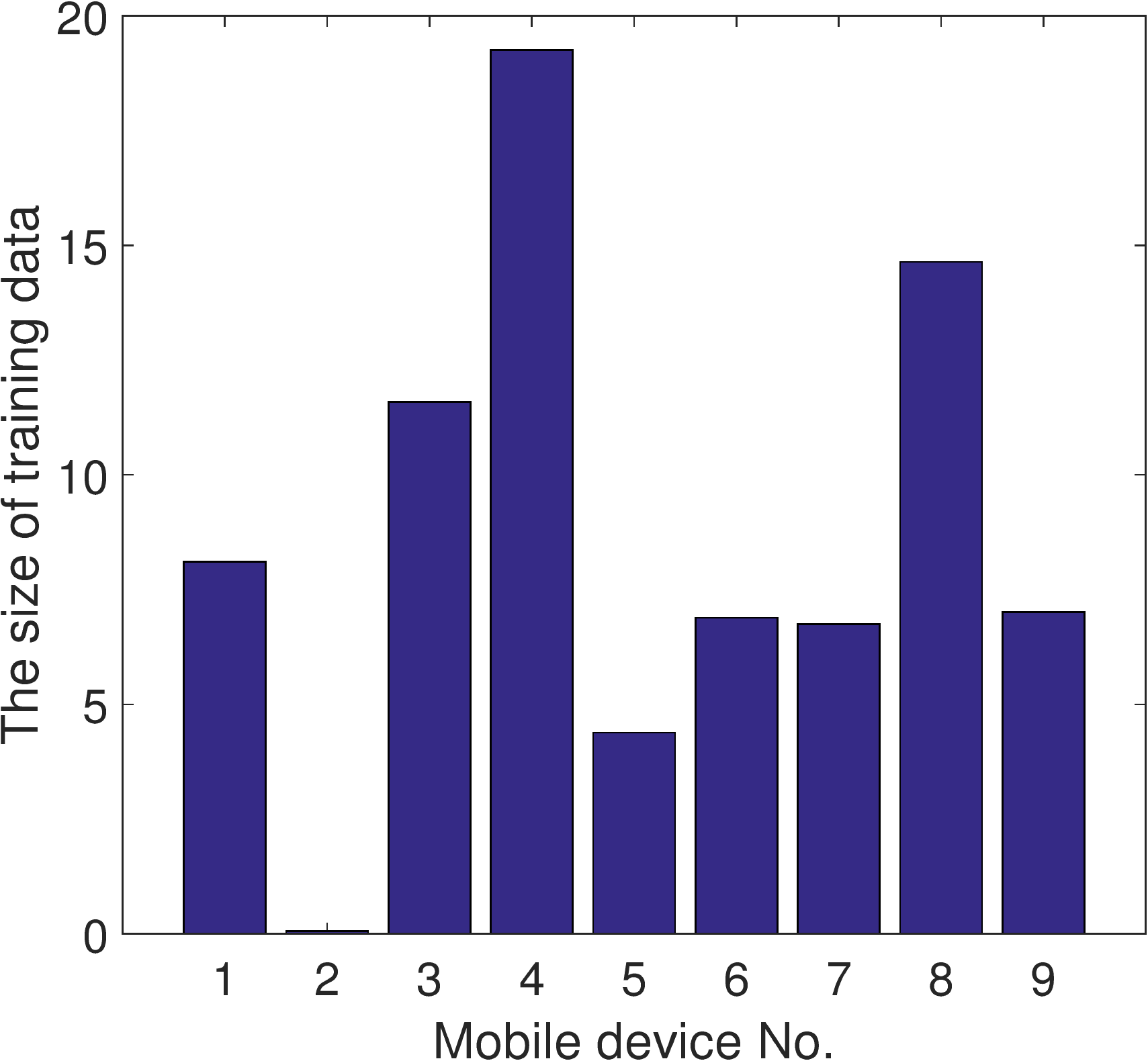}
 \caption{The size of training data for each mobile device.}
 \label{fig:demand}
\end{figure}

We then investigate the sizes of the training data for the model owner. As shown in Fig.~\ref{fig:demand}, the size of the training data from mobile device $4$ is the largest due to its lowest price as shown in Fig.~\ref{fig:price}. Along with the slow processing rate of mobile device $4$, i.e., $r^{\rm{p}}_4=61.65$, such a large size of data of mobile device $4$ implies that mobile device $4$ will take much time for performing computation on its training data. Accordingly, it is likely that the model updates from other mobile devices can arrive at mobile device $4$ before the mobile device $4$ finishes is its computation. Thus, the mobile device $4$ can serve as a relay for many other devices. As we observe in Fig.~\ref{fig:demand}, the sizes of data of mobile devices $5$ and $6$, i.e., the neighboring mobile devices of mobile device $4$ as shown in Fig.~\ref{fig:connection}, are much smaller than that of mobile device $4$. As a result, mobile devices $5$ and $6$ choose mobile device $4$ as their relay node. This helps mobile devices $5$ and $6$ to avoid transferring their model updates directly to the access point, improving the energy efficiency.

\section{Conclusion}
\label{sec:conclusion}

In this paper, we have presented the Stackelberg game model to analyze the transmission strategy and training data pricing strategy of the self-organized mobile device as well as the learning service subscription of the model owner in the cooperative federated learning system. We have focused on the interactions among the mobile devices and considered the impact of the interference cost on the mobile devices' profits. Moreover, we have investigated the impact of the size of the training data on both the model owner's utility and the mobile devices's profits. Specifically, we have established a model describing the impact of the mobile devices' transmission strategies on their transmission rates and relay node. The model also describes the impact of the model owner's learning service subscription strategy on the model owner's utility. We have studied the optimal strategy of the model owner by using best response and the equilibrium strategies of the mobile devices by using the exterior point method. Our future work will extend to the study in the long-run among the model owner and the mobile devices.


\end{document}